\documentstyle[12pt,aasms]{article}

\def\ts{{\thinspace}}
\def\lsim{\scriptstyle\buildrel < \over \sim}
\def\sgra{SGR\thinspace{0526$-$66}}
\def\sgrb{SGR\thinspace{1806$-$20}}
\def\sgrc{SGR\thinspace{1900$+$14}}
\def\g10{G\thinspace{10.0$-$0.3}}
\journalid{337}{15 January 1989}
\articleid{11}{14}

\slugcomment{Submitted to the Astrophysical Journal}

\begin{document}

\title{Radio Monitoring and High Resolution Imaging \\
of the Soft Gamma Ray Repeater {1806$-$20}}

\author{G. Vasisht}
\affil{California Institute of Technology, MS 105-24, Pasadena, CA 91125}

\author{D. A. Frail}
\affil{National Radio Astronomy Observatory, P.O. Box 0, Socorro, NM 87801}
\and
\author{S. R. Kulkarni}
\affil{California Institute of Technology, MS 105-24, Pasadena, CA 91125}




\begin{abstract}

We report the results of radio flux-monitoring and high resolution
observations at 3.6 cm with the VLA, of the central condensation in
\g10, the radio nebula surrounding the soft gamma ray repeater (SGR)
{1806}$-$20.
The quiescent X--ray source AX~1805.7$-$2025 is coincident with the
radio core suggesting that \g10\ is a plerionic supernova remnant.
The monitoring experiment was performed in 10 epochs
spread over five months, starting just before the latest reactivation
of \sgrb\ in $\gamma$-rays. There is no apparent increase in the radio
flux density from the central region of \g10\ on timescales of days to
months following the $\gamma$-ray bursts. At a resolution of 1 arcsec
the peak region of \g10\ is seen to consist of a compact source with
diffuse, one-sided emission, reminiscent of core-jet geometry seen in
AGNs and some accreting Galactic binaries.  By analogy with these
latter sources, we argue that the \sgrb\ is coincident with the core
component. If so, this is
the first arcsecond localization of a high energy transient. The lack
of radio variability and the low brightness temperature of the central
component distinguish \sgrb\ from other accreting binaries.
The structure of the high resolution radio image also
does not particularly resemble
that seen in the vicinity of young pulsars. Thus there is no compelling
observational evidence for either of the two models discussed for SGRs, the
pulsar
model or the accreting binary model.

\end{abstract}


%
\section{INTRODUCTION}

Since soft gamma ray repeaters  were first recognized as a separate
class of the gamma ray burster population with distinct temporal,
spectral and recurrence signatures (\cite{mg81},
\cite{Atteia87},
\cite{Kou87},
\cite{Lar87}), attempts to understand their nature have concentrated
primarily on their high energy emissions (\cite{Nor91}). The recent
suggestion by Kulkarni and Frail (1993) that \sgrb\ was
associated with the radio supernova remnant (SNR) G\ts{10.0$-$0.3}, and
the subsequent confirmation by the ASCA X-ray satellite (\cite{Mur94})
and the BATSE experiment (\cite{Kou94}) on board the Compton Gamma Ray
Observatory (CGRO), has underscored the importance of multi-frequency
observations for these objects.  Since then, a number of observations
have been reported including X-ray observations of \sgra\ and its SNR N49
(\cite{rkl94}), further detailed observations of \sgrb\ (\cite{Son94})
and  SNR G\ts{10.0$-$0.3} (\cite{Kul94a}), and searches for the SNR
counterpart to \sgrc\ (\cite{Kou94}, \cite{Hur94a}, \cite{Vas94}).

No model proposed thus far is commensurate with all the observations
but a general picture is starting to emerge from the various studies.
SGRs are young ($\lsim$10$^4$ yrs), energetic neutron stars.  The
rarity of SGRs (\cite{Kou94}, \cite{Hur94b})
means that either the SGR phase is short-lived phase
that all neutron stars undergo or that a small subset of the neutron
star population with some distinguishing characteristic evolve to SGRs
(e.g.\ large B-fields, high velocities, a special type of binary
system etc.).

Specifically for \sgrb\, two models explaining the observations of
nebula G\ts{10.0$-$0.3} and the X-ray source AX\ts{1805.7$-$2025} have
been proposed (\cite{Kul94a}). \begin{enumerate} \item [(1)] A radio
pulsar whose spin-down luminosity powers a synchrotron nebula or a
``plerion'' e.g. the Crab Nebula or 3C58. \item [(2)] An accreting
binary neutron star system with a radio nebula such as the one
associated with the X-ray binary Cir X-1 (\cite{Ste93}) or perhaps even
SS~433 (\cite{mar84}).  \end{enumerate}

Radio and X--ray observations offer the possibility of distinguishing
between these alternatives. In particular, the central sources in Cir
X-1 and SS 433 are easily detectable bright radio sources which undergo
dramatic changes in their flux density,
both in radio and X--ray bands.
In contrast, most pulsars are
faint radio sources with little changes in their flux density,
especially at high frequencies.

\sgrb\ is the most prolific of the three known SGRs, with greater than
$10^2$ burst events detected in several different instruments by the late
1970's and early 1980's (\cite{Ulm93}, \cite{Lar87}). Recently
\sgrb\ became active in $\gamma$-rays again after being quiescent for
nearly a decade (\cite{Kou94}). A total of six new bursts were
detected, including three on 29 Sept. 1993 and one each on 5 Oct., 9
Oct. and 10 Nov. 1993.  Fortunately, motivated by our earlier paper
(\cite{kf93}) we had begun and carried out extensive radio
observations of \g10, two weeks prior to the first BATSE burst
(\cite{Kul94a}).  Following the announcement of the 29 Sept. event we
initiated a radio monitoring program with the Very Large Array (VLA) to
search for variable radio emission from the radio counterpart of
\sgrb.   We also carried out high resolution imaging of the central
core of \g10\ in
an effort to accurately localize the stellar counterpart of
\sgrb.  In this {\it Letter} we report on both these observations.

\section{OBSERVATIONS AND RESULTS}

The observations were obtained at 3.6-cm with the VLA in its so-called
``D'' configuration (the most compact configuration).  In all there are
10 epochs of observations spanning nearly 5 months during which
there were a total of 6 $\gamma$-ray bursts (Table 1).
The pre-reactivation radio observation on 13 Sept.\ 1993 was
obtained in the ``DnC'' configuration, not too dissimilar to the
remaining observations.  As this observation was just prior to the
initial burst it provides a useful dataset against which to search for
changes in the radio emission from the source.

During each observation two separate IF bands, each measuring both
hands of circular polarization and each with a bandwidth of 50 MHz
were used at 8.415 GHz and 8.465 GHz (the 3.6-cm or X band).
Calibration of the data
and  reduction was done using the AIPS package. The total
integration time of the source depended on the epoch of observation,
and varied between 0.2 hrs and 1 hr.

The data from each epoch were added to form a single image and this is
shown in Figure 1. This image is similar but considerably deeper  than
the image from Kulkarni et al. (1994). A point-like ``core'' sits near
the center of a 1 arcmin nebulosity with extensions leading off to the
north and southwest. Since the aim of the experiment was to measure
flux variations across the source, especially in the core, images from
the various epochs were convolved to the same beam. These core fluxes
and total fluxes are shown in Table 1. Shown also is the time interval
between the epoch of each radio observation and the last $\gamma$-ray
outburst from \sgrb\ (see section 1; \cite{Kou94}).

We saw no significant variability for any of the components of
G\ts{10.0$-$0.3} during the 5 month interval in which we monitored this
source.  There are apparent variations at the 25\% level. These could
arise from a number of reasons. Most importantly observations of a complex
source such as \g10\ in the ``snapshot'' (short-integration) mode are
sensitive to the actual spatial frequency ($u$-$v$) coverage.  For
example, the observations of 22 November 1993 suffered from a shortage
of low spatial frequency data points resulting in an underestimation of
the total flux. We conclude that the source is essentially constant
at a level of 25\%.

All the previous observations of this source were obtained with the VLA
in compact configurations. In order to better localize the seat of
\sgrb\ as well as to study the source structure of the peak
region we observed the
source at 3.6-cm, on 7 March 1994, with the VLA in its most extended
configuration, the  ``A'' array.
At the highest angular resolution (0.2 arcsec) the compact source is resolved
(Figure 2a), giving a source dimension of $4\times 10^{16}d_{10}$ cm
where the distance to \sgrb\ is $10d_{10}$ kpc.
In order to improve our surface brightness sensitivity we tapered the
data in the $u$-$v$ domain. The resulting picture has a lower angular
resolution and is shown in Figure 2b.  The source appears to be quite
asymmetric with extended emission to the east. This picture of a core-jet
geometry
is not too dissimilar to that seen in AGNs, especially weak AGNs.  If
the analogy is correct, then the eastern extended feature probably
represents the outflow of material from the compact core.

We now review the observational facts with a view towards establishing
the seat of \sgrb.  We proceed by a series of steps from $\gamma$-rays
to radio wavelengths. First, the quiescent X--ray source AX 1805.7$-$2025
detected by Murakami et al.\ts{(1994)}
was localized to \g10\ and
is coincident with the radio peak.  This morphology is
characteristic of plerions and for this reason it is well accepted that AX
1805.7$-$2025 represents the compact source which powers \g10\ and is
also the X--ray counterpart of \sgrb.  The best localization of
AX~1805.7$-$2025 is by ROSAT (\cite{Coo93}) and this coincides with the
radio peak within the 90\% error radius of 11-arcsec.  Second, the
sequence of Figures 1, 2b and 2a strongly suggest that this compact
source is located at the peak in Figure 2a.  By analogy to AGN we
interpret the core-jet appearance in terms of an outflow of particles
from the compact source. If this chain of reasoning is correct then the
compact radio source is the seat of \sgrb.  

In view of the complexity of the radio peak region,
the localization of the core
is best done in the image with the highest angular resolution, Figure
2b.  This peak encompasses an integrated flux density of 0.6 mJy and is
centered at $\alpha$= 18$^h$ 05$^m$ 41.675$^s$ ($\pm{0.005}$),
$-20^\circ$ 25$^\prime$ 12.5$^{\prime\prime}$ ($\pm{0.1}$) (B1950).  We
henceforth refer to this component as the compact component, ``C''.
Subject to the caveats discussed above, this localization is  the first
subarcsecond localization of a high energy  transient.
More importantly
it allows deep searches to be made at optical and infrared wavelengths
and preliminary observations are reported in the companion paper
(\cite{Kul93b},\cite{Kul94b}).


Parenthetically we note that \cite{Kul94a}, using low resolution data
similar to Figure 1, reported a position for \sgrb\ which is offset
from that of ``C'' in Figure 2. by 1.3 arcsec.  The origin of the
offset is clearly due to the asymmetric nature of the source, not fully
resolved in the low resolution image reported in \cite{Kul94a}.

\section{DISCUSSION}

Despite an extensive monitoring campaign of \g10\ with post-burst
timescales ranging from as little as two days after a $\gamma$-ray burst
to three months after the bursting finally ceased, we detected no radio
variability above the 25\% level. Variability at this level can be
reasonably attributed to uncertainties due to calibration and varying
spatial frequency coverage (section 2). Thus for all practical purpose,
the radio counterpart of \sgrb\ is constant.  Indeed, the constancy of
both the radio and X-ray flux (\cite{Son94}) seems to be one of the
defining characteristics of the emission surrounding \sgrb, punctuated
only by brief (0.1 sec), intense ($\sim$ 10$^{41}$ erg s$^{-1}$) $\gamma$-ray
bursts.

Dramatic radio flaring is often seen in Galactic binaries such as Cyg
X-3, Cir X-1 and SS\ts{433}, which are correlated with episodes of
X-ray activity in these sources (\cite{sea94}, \cite{hje88}).  VLBI and
VLA studies of these sources (Cir X-1, \cite{Pre83}; Cyg X-3,
\cite{Gel83}; SS 433, \cite{hj81})
reveal bright ($\sim$ Jy or greater), highly variable compact
component (sizes typically in the range of a milli-arcsec to tens of
milli-arcsecs) coincident with the stellar binary source.
Most likely a
fraction of the gravitational potential energy of the accreted material
is harnessed to accelerate particles to relativistic energies primarily
during episodes of super-Eddington accretion (\cite{Ino89}). The
implied
brightness temperatures are in excess of $10^6$ K at centimeter
wavelengths.  In contrast, component ``C'' in \sgrb\ is dim, only
about 0.6 mJy (section 2) and is resolved by a beam of 0.2 arcsec. Here
we make the reasonable assumption that, given the high frequency of the
observations, the observed size is intrinsic  and not a result of
broadening by interstellar scattering.

Thus the compact component in \sgrb\ differs from the radio
counterpart of
accreting neutron star systems associated with radio nebulae in two
substantial ways:  ({\it i}) no dramatic variation in the total flux has been
seen and ({\it ii}) the brightness temperature is small, about 100 K. We
mention a caveat with reference to ({\it i}) above.  The peak flux values in
Table 1 refer to the entire core region, of which 70\% is due to the
extended emission in Figure 2.  Since the total flux density of the
core region did not change significantly
($<\hskip-2pt{\rm{S_p}}\hskip2pt>$$\sim$4.8$\pm$1.0 mJy) either before,
during or after the $\gamma$-ray bursts, it seems that the one-sided
outflow, if it originated from during the last bursting phase of \sgrb,
did not produce dramatic synchrotron flaring.  However, for variations in the
integrated flux in ``C''  to have been discerned above
the extended emission from the core it would had to increase in flux
density by more than a factor of 10.  Thus we cannot rule out
variability in ``C'' at the factor of 10 level. Nevertheless, this
is in marked contrast to other radio variables including Cyg X-3 and
LSI+61$^\circ$ 303 whose flux densities during the flaring phase rise
to 10$^2-10^4$ of their quiescent values (Hjellming 1988).

We note that the size of ``C'' which, as argued
above, appears to be quiescent is not very different from that in
Cir X-1 (0.2 arcsec; \cite{Pre83}) and GRS 1915+10 (0.1 arcsec;
\cite{mr94}). The implied size, $4\times 10^{16}$ cm
is much larger than  any binary orbit. Thus, as in other radio
emitting neutron star binaries,
the energetic particles are injected on scales much larger than
either the neutron star or any potential binary orbit.

In what way do the new observations constrain possible models for SGRs?
The constancy of the radio and X--ray emission are in accord with the
pulsar model.
Structure in radio emission is often seen around
energetic pulsars and
usually interpreted to be the working surface of the pulsar
relativistic wind as it shocks the nebular gas e.g.  the wisps in the
Crab nebula or the bow shock seen around several pulsars.
However, it is fair to say that the one-sided structure
seen in Figure 2b is quite unusual for a pulsar powered plerion.

Some authors have suggested a connection between fast moving
neutron stars and SGRs, both to explain the rarity of SGRs and
the greatly offset position of SGR~0526$-$66 from the
center of the SNR N49 (\cite{Cli82}). If so, the asymmetric
structure seen in Figure 2b can be interpreted as a bow shock
of a rapidly moving pulsar. However, we consider the bow shock hypothesis
to be unlikely. Specifically we draw the reader's attention to the
fact that the east-west asymmetry is seen on angular
scales ranging from 1-arcmin (Figure 1) to 1-arcsec (Figure 2b).
If this is the wake of an
ejected pulsar then the wake should be directed radially outwards
with respect to the center which is not the case because the SGR is
offset to the NE with respect to the center of G\ts10.0$-$0.3. We
suggest instead that there has been a
one-sided outflow from this object for quite some time. This hypothesis
(see below) satisfactorily explains the observations that the
east-west asymmetry is seen on large and small scales.

The lack of radio flaring, the low brightness temperature and the
lack of a marked flow geometry make the central compact
component quite unlike
the radio counterparts of accreting neutron star binaries.
Other difficulties with the binary model
have been raised by Sonobe et al. (1994), including the constancy of
the quiescent X-ray emission and its low luminosity
(L$_x\simeq{10}^{35}$ erg s$^{-1}$).  While we cannot rule out the
binary model due to our limited number of radio observations, it is
reasonable to conclude that the existing data do not necessarily
favor a binary model.

Finally we come to the important issue of the role played by the
outbursts in powering the plerion. Kulkarni et al. (1994) have
suggested that the
$\gamma$-ray outbursts are accompanied by an energetic (but an
otherwise invisible) particle outburst and these outbursts cumulatively
power the nebula.  Our observations allow us to derive an upper limit
to the ratio of the particle luminosity to photon luminosity, $\beta$.
The 2-arcsec feature in Figure 2b corresponds to $4d_{10}$ light
months, comparable to the duration of our radio monitoring. We apply
the standard synchrotron model (\cite{Pac70}) to this feature to deduce
$E_{min}$, the minimum energy in particles and magnetic fields.
Assuming low and high cutoffs of 10 MHz and 100 GHz for radio
luminosity, we can make an estimate of the minimum energy in particles
and magnetic fields. Owing to the lack of direct spectral information we assume
that these structures have a spectral index same as that of the entire
nebula ($\alpha$ $\sim$ $-$0.6). We estimate $E_{min} \sim 1\times
10^{45}d_{10}^{17/7}$ ergs. Over this period, a total of six
bursts with a mean fluence of $10^{-7}$ erg cm$^{-2}$ have been
detected by BATSE (\cite{Kou94}). Thus the total photon burst
energy is $10^{40}d_{10}^2$ erg.

The fact that we see no change in the radio emission of the central
region sets an upper bound to the particle energy to be less than
$\sim 10$\ts$E_{min}$ (see above).
Thus $\beta$ $\lsim$ 10$^6$, roughly independent of distance. This is
a true upper limit since we have assumed the propagation speed for
particles to be equal to the speed of light. Unfortunately, this limit
is not very interesting since one naively expects $\beta$ to be around
unity rather than such a high number. However, this simple exercise
demonstrates that in principle $\beta$ can be constrained by
observations alone. Clearly, it is of great value to continue high
angular resolutions of the compact core.

We conclude that, with the available data, it is not possible
to assert whether the bursts power the nebula.  Regardless of this
we note that the same east-west asymmetry exists on scales
of arcseconds to arcminutes which suggests that the on-going energy
injection, impulsive or steady, has been occuring over a long time
scale.

\bigskip

\noindent{{\bf Acknowledgements:} We thank M. Rupen for his
assistance with the October observations and DAF acknowledges useful
discussions with R. Hjellming. The VLA is is operated by Associated
Universities Inc. under cooperative agreement with the US National
Science foundation.}

%
%

%

\clearpage

\begin{figure}

\noindent Fig. 1.--- A 3.6-cm radio continuum image of a
one arcmin region around the
central core of G\ts{10.0$-$0.3}. The image was made using co-added
datasets from the various monitoring epochs. Extensions are seen
prominently to the East and also the North and South-West to a
lesser extent. Contours are at 90, 360, 450, 540, 630, 720, 810,
900, 990, 1080, 1170, 1260, 1350, 1440, 1530, 1620, 1710 and 1800
$\mu$Jy. The rms noise is 40 $\mu$Jy/beam and the peak flux is
1.8 mJy. The synthesised beam is 6.6 arcsec by 4.9 arcsec at a
position angle of $-$86 degrees (measured counter clockwise).
\bigskip

\noindent Fig. 2a.--- A 3.6-cm radio continuum image of the radio ``core'' of
G\ts{10.0$-$0.3}, revealing the arcsecond core which we postulate is
the seat of the SGR ($\alpha$= 18$^h$ 05$^m$ 41.675$^s$ ($\pm{0.005}$),
$-20^\circ$ 25$^\prime$ 12.5$^{\prime\prime}$ ($\pm{0.1}$) (B1950)).
Faint structure is seen to the East of the
central source. The image was made with natural weighting, producing
a synthesized beam of 0.45 arcsec by 0.26 arcsec and position angle
(measured counter clockwise) of 12 degrees. Contours are $-$60, 70, 80,
90, 100, 110, 120, 130, 140 and 150 $\mu$Jy/beam. The rms noise is 20
$\mu$Jy/beam and the peak flux density is 0.15 mJy.
\bigskip

\noindent Fig. 2b.--- A 3.6-cm radio continuum image of radio ``core'' of
G\ts{10.0$-$0.3}, suggesting a collimated flow to the East
of the central source, on scales of 2 arcsec. The image was made by
uniform weighting of the data, along with a taper which resulted
in a synthesised beam of 0.9 arcsec by 0.8 arcsec. This procedure
is sensitive to picking up extended flux. Contours are at $-$90, 90,
150, 210, 240, 270, 300, 330 $\mu$Jy. The rms noise is 30 $\mu$Jy/beam
and the peak flux density is 0.3 mJy.

\end{figure}

%



\begin{thebibliography}{}

\bibitem[Atteia et al. 1987] {Atteia87}\reference Atteia, J. L. et al. 1987,
              \apj, 320, L105 (1987)

\bibitem[Cline et al. 1982]{Cli82} Cline, T. L. et al. 1982,
                 \apj, 255, L45

\bibitem[Cooke 1993]{Coo93} \reference Cooke, B. A. 1993, Nature, 366,
413

\bibitem[Hjellming \& Johnston 1981]{hj81} \reference
Hjellming, R. M. \& Johnston, K. J. 1981, Nature, 290, 100

\bibitem[Hjellming 1988]{hje88} \reference Hjellming, R. M. 1988, in
Galactic and Extragalactic Radio Astronomy, 2nd Edition, Eds. G. L.
Verschuur and K. I. Kellermann (Springer-Verlag: New York), p 381

\bibitem[Hurley et al. 1994]{Hur94a} \reference Hurley, K., Sommer, M.,
Kouveliotou, C., Fishman, G., Meegan, C., Cline, T., Boer, M. \& Niel,
M. 1994, ApJ, 431, L31

\bibitem[Hurley et al. 1994]{Hur94b} \reference Hurley, K.,
Atteia, J. L, Jourdain, E., Barat, C., Niel, M., Vedrenne, G.,
Sunyaev, R., Kuznetsov, A. and Terekhov, O. 1994,
\apj, 423, 709

\bibitem[Geldzahler et al. 1983]{Gel83} \reference Geldzahler, B. J.
et al. 1983, \apj, 273, L65

\bibitem[Inoue 1989]{Ino89} \reference Inoue, H. 1989,
Proc. 23rd ESLAB Symp. on Two-Topics in X-ray Astronomy, ESA SP-296,
109

\bibitem[Kouvelioutou et al. 1994]{Kou94} \reference Kouvelioutou, C.
et al. 1994, Nature, 38, 125

\bibitem[Kouvelioutou et al. 1987]{Kou87} \reference Kouvelioutou, C.
et al. 1987, \apj, 322, L21

\bibitem[Kulkarni \& Frail 1993]{kf93} \reference Kulkarni, S. R. \&
Frail, D. A. 1993, Nature, 365, 33


\bibitem[Kulkarni et al. 1993b]{Kul93b}\reference Kulkarni, S. R.,
Matthews, K., Neugebauer, G., Reid, N., Soifer, T. \& Vasisht, G.
1993, IAU Circ. 5883

\bibitem[Kulkarni et al. 1994]{Kul94a} \reference Kulkarni, S. R.,
Frail, D. A., Kassim, N. E., Murakami, T. \& Vasisht, G. 1994, Nature,
368, 129

\bibitem[Kulkarni et al. 1994b]{Kul94b}\reference Kulkarni, S. R.,
Matthews, K., Neugebauer, G., Reid, N.,van Kerkwijk, M. H. \& Vasisht, G.
1994, \apj, submitted

\bibitem[Laros et al. 1987]{Lar87} \reference Laros, J. G. et al.
1987, \apj, 320, L111

\bibitem[Margon 1984]{mar84}\reference
  Margon, B. 1984, Annu. Rev. A\&A, 22, 507

\bibitem[Mazets \& Golenskii 1981]{mg81} \reference Mazets, E. P.
\& Golenskii, S. V. 1981, Astrophys. Space Sci, 75, 47

\bibitem[Mirabel \& Rod\'riguez 1994]{mr94} \reference Mirabel, I. F. \&
Rod\'riguez, L. F. 1994, Nature, 371, 46


\bibitem[Murakami et al. 1994]{Mur94} \reference Murakami, T. et al.
1994, Nature, 368. 127

\bibitem[Norris et al. 1991]{Nor91} \reference Norris, J. P., Hertz,
P., Wood, K. S. \& Kouvelioutou, C. 1991, \apj, 366, 240

\bibitem[Pacholczyk 1970]{Pac70} \reference Pacholczyk, A. G. 1970, Radio
Astrophysics, Chap. 7, San Fransisco: Freeman


\bibitem[Preston et al. 1983]{Pre83}\reference Preston, R. A. et al.
1983, ApJ, 268, L23

\bibitem[Rothschild, Kulkarni, \& Lingenfelter 1994]{rkl94} \reference
Rothschild, R. E., Kulkarni, S. R., \& Lingenfelter, R. E. 1994,
Nature, 368, 432

\bibitem[Seaquist 1994]{sea94} \reference Seaquist, E. R. 1994,
Rep. Prog. Phys., 56, 1145

\bibitem[Sonobe et al. 1994]{Son94} \reference Sonobe, T., Murakami,
T., Kulkarni, S. R., Aoki, T. \& Yoshida, A. 1994, \apj, (in press)

\bibitem[Stewart et al. 1993]{Ste93} \reference Stewart, R. T.,
Caswell, J. L., Haynes, R. F. \& Nelson, G. J. 1993, MNRAS, 261, 593

\bibitem[Ulmer et al. 1993]{Ulm93} \reference Ulmer, A., Fenimore, E. E.,
Epstein,
R. I., Ho, C \& Klebesadel, R. W. 1993, \apj, 418, 395

\bibitem[Vasisht et al. 1994]{Vas94} \reference Vasisht, G, Kulkarni,
S. R., Frail, D. A. \& Greiner, J. 1994, \apj, 431, L35

\end{thebibliography}
\end{document}